\documentstyle[11pt,aaspp4]{article}
 
\slugcomment{Accepted by ApJ for Feb 2001 issue}

\lefthead{Leggett et al.}
\righthead{Spectral Energy Distributions of L--Dwarfs}
 
\begin{document}
 
\title{Infrared Spectra and Spectral Energy Distributions of Late--M--
and L--Dwarfs}

\author{S.K. Leggett}
\affil{Joint Astronomy Centre, University Park, Hilo, HI 96720
\nl skl@jach.hawaii.edu}
 
\author{F. Allard} 
\affil{CRAL, Ecole Normale Superieure, 46 Allee d'Italie, Lyon, 69364 France
\nl  fallard@cral.ens-lyon.fr } 

\author{T.R. Geballe}
\affil{Gemini Observatory, University Park, Hilo, HI 96720
\nl tgeballe@gemini.edu}

\author{P.H. Hauschildt}
\affil{Department of Physics and Astronomy \& Center for Simulational Physics, 
\nl University of Georgia, Athens, GA 30602-2451
\nl yeti@hal.physast.uga.edu}

\and 
\author{Andreas Schweitzer}
\affil{Department of Physics and Astronomy \& Center for Simulational Physics, 
\nl University of Georgia, Athens, GA 30602-2451
\nl andy@physast.uga.edu}

\begin{abstract} We have obtained 1.0---2.5$\mu$m spectra at R$\sim$600 of
14 disk dwarfs with spectral types M6 to L7. For four of the dwarfs we 
have also obtained infrared spectra at R$\sim$3000 in several narrow
intervals. In addition, we present new L$^{\prime}$ photometry for four of the
dwarfs in the sample, which allows improved determinations of their
bolometric luminosities.  While obtaining the photometry we resolved
the L-dwarf Denis-P J 0205-1159 into an identical pair of objects separated
by 0\farcs 35. The spectra, together with the published energy
distribution for one other L5 dwarf, are compared to synthetic spectra
generated by upgraded model atmospheres.  Good matches are
found for 2200$\ge$$T_{\rm eff}$~K$\ge$1900 (spectral types around M9 to
L3), but discrepancies exist at $T_{\rm eff}\geq$2300~K (M8) and for
$T_{\rm eff}\leq$1800~K (L4--L7).  At the higher temperatures the
mismatches are due to incompleteness in the water vapor opacity
linelist.  At the lower temperatures the disagreement is probably due
to our treatment of dust: we assume a photospheric distribution in
equilibrium with the gas phase and neglect any diffusion mechanisms.
We derive effective temperatures for the sample from the comparison
with synthetic spectra and also by comparing our observed
total intrinsic luminosities to structural model calculations
(which are mostly independent of the atmosphere but are dependent on
the unknown masses and ages of the targets).  The two derivations
agree to about 200~K except for the faintest object in the sample
where the discrepancy is larger.  Agreement with other temperature
determinations is also $\pm\sim$200~K, except for the L7 dwarf.

\end{abstract}
 
\section{Introduction}

The last few years have seen rapid advances in both the observational data
on low--mass stars and brown dwarfs and in the theoretical models of
these objects. During this period we have been obtaining infrared spectra of
a sample of halo and disk dwarfs approaching and below the
stellar/sub--stellar boundary.  This paper extends our spectroscopic study
of M dwarfs presented in \cite{l99} (hereafter L00) to lower effective
temperatures ($T_{\rm eff} <$ 2500~K). The data are compared to synthetic
spectra generated from an upgraded version of Allard \& Hauschildt's
NextGen model atmospheres (\cite{all98b}).

In \S 2 we present the target sample and in \S 3 we describe the
instruments used to obtain the spectroscopic data. In \S 4 we present
imaging results:  new
L$^{\prime}$ photometry, confirmation of binarity of Denis-P J
0205-1159, and color--color diagrams. In \S
5 we describe the spectra and calculate integrated fluxes and bolometric
magnitudes. \S 6 describes the models and the comparison process and the
results of the comparison of the data to the synthetic spectra are
presented in \S 7.  In \S 8 we present derivations of effective temperature
based on structural models. Our conclusions are given in \S 9.

\section{The Sample}

The fourteen dwarfs whose spectra are reported here were selected from
various color or proper--motion surveys --- \cite{d97}
(the DENIS-P survey, sometimes known as DBD objects); \cite{i91} (the BRI 
objects); \cite{k96}
(the 2MASS survey);  \cite{rla97} (Kelu-1);  \cite{t93} (the TVLM objects)
--- or from follow--up studies of known high proper--motion objects ---
\cite{k97}; \cite{m92} --- or from studies of low--mass companions to
nearby stars --- \cite{bz88} (GD 165B).  We also include for analysis the
previously published energy distribution for the L dwarf SDSS~0539 found by 
the Sloan Digital Sky Survey  (\cite{f00,l00}). The targets are
listed in Table 1. We  give discovery name, or LHS or LP number
(\cite{luy79}), and/or Gliese or Gliese/Jahreiss number (\cite{gj91}), for
each dwarf.  An abbreviated RA/Dec is also given to aid identification.  
The infrared instrumentation used for each object is given, using the
configuration names from Table 2 (see \S 3).

The spectral types in Table 1 are based on optical spectra and are taken
from Kirkpatrick et al. 1995, 1999a, 2000, \cite{k97} and \cite{f00}.
These classifications are all on the same system; we note the
significantly earlier spectral type assigned to Denis-P J 0205-1159 
according to a classification system by \cite{mar99b}.  The differences
between the two classification systems are discussed further in \S 5.1. 

The kinematic populations have been taken from \cite{l98,l92},
except for Kelu-1, Denis-P J 0205-1159, Denis-P J 1058-1548 and Denis-P J 1228-1547 for
which we have calculated space velocities using parallaxes and proper
motions from \cite{nofs99} and radial velocities from \cite{bas97}, 2000.
The space velocities of these four objects are low (U 20 to 30 km/s, 
V $-13$ to $-20$ km/s, W 1 to 5 km/s) and imply young disk population. The
kinematic classification schemes for young disk (YD), young/old disk (Y/O) 
and old disk (OD) are described in \cite{l92}.

Detections of lithium in Kelu-1 (\cite{rla97,bas97}), Denis-P J 1228-1547
(\cite{bas97,t97}) and LP 944-20 (\cite{t98a}) imply that each of these
objects is substellar and at an age of $\sim$1~Gyr.  A non--detection for
Denis-P J 1058-1548 (\cite{bas97}) implies a substellar mass if its age is
$\sim$1~Gyr or a mass at the stellar/substellar transition of
0.075~M$_{\odot}$ if it is several Gyr.  The age of GD 165B also is
uncertain; its mass is near the stellar/substellar boundary, but it is
likely to be a brown dwarf (\cite{k99b}).

Denis-P J 1228-1547 was recently resolved to be a double system by \cite{mar99},
using the Hubble Space Telescope.   The binarity of Denis-P J0205-1159 was 
announced by \cite{koe99} on the basis of K--band observations at Keck;
in \S4.2 we confirm this using JHKL$^{\prime}$ imaging data from the 
United Kingdom 3.8~m Infrared Telescope (UKIRT).   2MASP J0345432$+$254023
is suspected to be a double--lined spectroscopic binary
(\cite{rei99}).  Indirect evidence for further multiplicity in
our sample is discussed in \S 4.3. 

\section{Spectroscopic Observations and Data Reduction}

The infrared spectra were obtained at UKIRT and at the University of
Hawaii 2.2~m telescope, using
some of the same spectrographs as L00.  Table 2 gives the
observing dates, grating and slit information, and resolutions. The
near--infrared data for SDSS~0539 were obtained using the configuration
designated IR1 in Table 1, except that the slit was wider (1$\farcs$2),
yielding resolutions of 25~A at J and 50~A at H and K. The lower
resolution data cover the entire JHK bandpasses, from 1.0$\mu$m to
2.5$\mu$m. The medium resolution data cover narrower spectral regions:
1.19---1.30$\mu$m, 1.58---1.67$\mu$m, and 2.02---2.18$\mu$m. The data are
available in electronic form on request.

Data reduction was carried out in the usual way using the Figaro software
package.  The effect of the terrestrial atmosphere was removed by dividing
by the spectra of nearby early--type stars, after removing hydrogen
lines seen in their spectra. The shape of each infrared spectrum
was corrected for the known flux distribution of the early--type star.  
The low--resolution spectral segments were individually flux--calibrated
using previously reported or recently obtained JHK photometry. Each
segment was integrated over the appropriate filter profile (UKIRT JHK) and
the observed flux from Vega was integrated over the same profile;  Vega
was assumed to be zero magnitude at all wavelengths, and the target flux
was scaled to match the broadband photometry.  The higher resolution data
were scaled to match the calibrated lower resolution data.

Optical spectra were obtained from the literature for all the objects in
the sample; the data sources are listed in Table 1.  These data were
flux-calibrated 
using the Cousins I filter and available I photometry, except for SDSS
0539 which was flux--calibrated using Sloan Digital Sky Survey photometry
(\cite{f00}).

\section{Imaging Results}

\subsection{New Photometry}

New infrared photometry was obtained for four of the objects in the
sample. In 1999
February and March L$^{\prime}$ data were obtained for Denis-P J 1058-1548 and
Denis-P J 1228-1547AB using IRCAM3 (pixel scale 0.28 arcseconds/pixel) at UKIRT.  
In 1999 September data were obtained for Denis-P J 0205-1159 (JHK and
L$^{\prime}$) and 2MASP J0345432$+$254023 (L$^{\prime}$ only), using the same
camera, which had been modified to a smaller pixel scale of
0.08 arcseconds/pixel (hereafter IRCAM/TUFTI).  The results are
presented in Table 3.

Standards with a range of color were observed; the results indicated that
no color term was needed to convert the L$^{\prime}$ data taken with
IRCAM/TUFTI to the established UKIRT (IRCAM3) L$^{\prime}$ system.  
However the J, H and K photometric systems are different due to significant
differences in the filter set.  The results in Table 3 are given in the
new photometric system which we refer to here as the ``UKIRT--UFTI(MKO--NIR)''
system as the UFTI camera, with the new MKO--NIR filter set, is being used to
define this system.  Both UKIRT cameras, UFTI and IRCAM/TUFTI, are now
configured with the new Mauna Kea consortium JHKL$^{\prime}$ filter set.  
Transformations between the established UKIRT system and the new MKO--NIR
system are presented in \cite{haw00}.

\subsection{Denis-P J 0205-1159 : an Identical Pair of L--Dwarfs}

Denis-P J 0205-1159, which was observed in excellent seeing conditions on 1999
September 19 (UT) using the IRCAM/TUFTI camera, was easily resolved as a
double system separated by 0.35$\pm$0.03 arcseconds at a position angle of
77$\pm$4 degrees. Figure 1 shows the K image of the system.  The
JHKL$^{\prime}$ colors of each component  are identical to
within the measurement error. Proper motion measurements by \cite{koe99}
show that the object is a true binary. Those authors obtain a separation
of 0.51$\pm$0.02 arcseconds and a position angle in 1997 July of 106$\pm$5
degrees and, in 1999 January, 72$\pm$10 degrees.  We suggest that our
values of 0.35$\pm$0.03 arcseconds and 77$\pm$4 degrees, based on higher
resolution data (0.08 arcseconds/pixel cf. 0.15 arcseconds/pixel), are more 
accurate.  The 0$\farcs$35 separation translates
to a physical separation of 6.3~AU.  This system should be monitored to
obtain orbital information.

\subsection{Color--Color Diagrams}

Table 4 lists distance moduli and VIJHKL$^{\prime}$ colors for the sample.
The V and I photometry were obtained from \cite{nofs99} for Denis-P J
0205-1159AB, Denis-P J 1058-1548, Denis-P J 1228-1547AB, 2MASP J0345432$+$254023 
and Kelu-1.  The I
magnitude for SDSS 0539 is synthesized from the flux--calibrated spectrum.  
J, H and K magnitudes for this object are from \cite{l00}.  The L$^{\prime}$ 
magnitude for GD 165B
is taken from \cite{jon96}.  Otherwise the photometry is from this work or
taken from the compilations by \cite{l92,l98}.  The infrared photometry is
on the established UKIRT (IRCAM3) system.  We have averaged the JHK values
for Denis-P J 0205-1159AB presented here with the earlier values in \cite{l98}
after transforming the new JHK to the old photometric system using the
known filter profiles.  The two sets of values agree well.

Figures 2--4 show various VIJKL$^{\prime}$ color--color diagrams; although 
similar diagrams are shown in \cite{l98} the ones shown here use improved or
additional data at V, I and L$^{\prime}$, and are useful references
as clear trends are seen.  Object labels in these and 
following plots are abbreviated: Denis-P J0205-1159AB, Denis-P J 1058-1548, 
Denis-P J 1228-1547AB become DNS 0205, 1058 and 1228 respectively;
2MASP J0345432$+$254023 is labelled as 2M 0345; TVLM 513-46546 is shown as
T 513-46546.

Figure 3 shows a synthetic color--temperature sequence for
solar--metallicity 
and $\log(g)=$5.0, generated from Ames--Dusty--2000 models (see \S 6) by 
convolving the calculated energy distributions with the known UKIRT filter
profiles.  There are some discrepancies between
the calculated and observed colors due to known inadequacies in the
models which we discuss in \S 6 and \S 7.  Nevertheless this comparison
provides the temperature range required for the spectral analysis
model grid (approximately 1800~K to 2300~K). 

Figure 5 shows M$_J$:J$-$K with isochrones from \cite{cbah00} for ages
5~Gyr and 0.1~Gyr. These isochrones use an earlier very similar Dusty
version of the NextGen model atmospheres. The agreement is poor for
the hotter dwarfs with M$_J<$12, and not good for the coolest objects 
Denis-P J 0205-1159 and Denis-P J 1228-1547. The discrepancies reflect 
known inadequacies in the atmospheric models used to generate the colors 
(\S 6) rather than problems with the structural models.  

As mentioned in \S 3, 
LP 944-20, Kelu-1 and Denis-P J 1228-1547AB are young substellar
objects, and GD 165B (L4) and Denis-P J 1058-1548 (L3) are substellar or
stellar/substellar transition mass objects, depending on their age.
Kelu-1 is superluminous but no companion is seen in HST images
(\cite{mar99}). \cite{mar99b} point out that if Kelu-1 is a single
object its luminosity implies an age of $<$0.1~Gyr. \cite{bas00} find
that the object is a very rapid rotator ($vsini=60$km/s) and so
it may indeed be very young, if the correlation seen for hotter low--mass 
stars between youth and rapid rotation holds here.  Nevertheless it is 
important to monitor it for radial velocity variations.  2MASP 
J0345432$+$254023, which also appears superluminous, 
is suspected to be a spectroscopic binary (see \S 2). Both Denis-P J 0205-1159
and Denis-P J 1228-1547 are known to consist of pairs of very similar
objects. The high frequency of binaries is not surprising in an
effectively magnitude--limited sample such as this.  Note that the evolutionary
models imply that each component of the $\sim$L5 dwarfs Denis-P J 0205-1159
and Denis-P J 1228-1547 must be substellar even for an age of several Gyr.

\section{Spectroscopic Results}

\subsection{Spectroscopic Sequences}

Figure 6 shows a representative set of R$\sim$600 spectra covering the
temperature range of our sample, which corresponds to spectral types dM7
to around dL7. The obvious features to note are: the CO bands at
2.3--2.4$\mu$m; the water bands at 1.4$\mu$m, 1.8$\mu$m and longward of
2.4$\mu$m;
the FeH bands at 0.99$\mu$m and 1.2$\mu$m; and the KI doublets near
1.18$\mu$m and 1.24$\mu$m.  Note that the TiO bands, which are used to 
classify the M dwarfs, are not apparent or very weak in the L dwarf spectra. 
A pseudo--continuum approach has been
used by \cite{mar99b} to classify the L dwarfs from optical spectra. They
find a good correlation between this scheme and the strength of the water
band at 1.4$\mu$m.  The trend of strengthening water absorption with
later spectral type is seen in Figure 6. \cite{k99} use ratios of various 
optical absorption features to classify the L dwarfs. Both schemes roughly 
agree up to L5 at which point the Mart\'{\i}n classes become earlier than 
Kirkpatrick's (so that, for example, Denis-P J 0205-1159 is classified L5 
in the Mart\'{\i}n scheme and L7 on Kirkpatrick's).

We note that the shape and depth of the water bands are calculated to
be strongly affected by the presence of dust, which affects the emergent 
spectra for $T_{\rm eff}\leq 2500$~K, or spectral types dM7 and later.
With no dust in the atmosphere, the bands are predicted to become deeper
with increasingly steep wings as temperature decreases.  When dust is present 
the atmosphere is heated in the line forming region and the bands become 
shallower and wider.  It is likely that as
infrared spectra are obtained for a larger sample of
L dwarfs, more scatter will be seen in the relationship between the depth of 
the water bands and effective temperature or (optically derived) 
spectral type, as dust
properties will vary from object to object (due to e.g. rotation,
metallicity or age differences).

\subsection{Integrated Fluxes and Bolometric Corrections}

Table 4 gives integrated fluxes for the sample, expressed as flux at the
Earth, apparent bolometric magnitude and intrinsic stellar luminosity.  They 
were obtained by integrating the infrared spectra over wavelength and adding
estimated flux contributions at shorter and longer wavelengths. Each
shorter wavelength contribution was approximated as a linear extrapolation
to zero flux at zero wavelength. The contributions at wavelengths beyond
2.4$\mu$m were calculated by deriving the fluxes at L$^{\prime}$ using an
effective wavelength approach, interpolating the spectrum from the end of
the K--band spectrum to the L$^{\prime}$ wavelength, and assuming a
Rayleigh--Jeans tail beyond L$^{\prime}$ (model calculations imply that the
Rayleigh--Jeans assumption is good to $\sim$1\% at these temperatures).
Although it is now known that methane absorption at 3.3$\mu$m is seen in
late L--dwarfs (\cite{noll00}), the feature is weak and will not
significantly effect the determination of total flux for this sample.
For SDSS 0539 we adopted an
L$^{\prime}$ magnitude based on J$-$K, the K magnitude, and Figure 4.  We
note that for dwarfs with 2700~K $\ge$ $T_{\rm eff} \ge$ 2000~K (spectral
types later than dM5.5)  the flux beyond 2.4$\mu$m makes up
$>$20\% of the total flux and hence L$^{\prime}$ measurements are
crucial for an accurate determination of luminosity. The uncertainties are
5\%, 0.05~mag and 0.02~dex in total flux, bolometric correction and
log$_{10}L/L_{\odot}$, respectively.

Figure 7 plots the K--band bolometric correction against I$-$K and Figure 8
shows BC$_K$ against J$-$K. We have
included the results from \cite{l96} and L00.  For the non--halo (disk and
likely disk) dwarfs only, the relationship between K--band bolometric
correction and color is represented by the cubic polynomials:
$$ BC_K =  1.307 +  0.9147(I-K) - 0.1592(I-K)^2 + 0.01054(I-K)^3 $$
$$ BC_K =  -0.31 +  5.124(J-K) - 2.031(J-K)^2 +  0.13877(J-K)^3 $$
for 1.9 $\leq$ I$-$K $\leq$ 5.5 and 0.75 $\leq$ J$-$K $\leq$ 1.60. These 
relationships are indicated by the 
solid lines in Figures 7 and 8, and supercede that given in L00.

\section{Models and Synthetic Spectra, and Comparison Process}

The  Ames--Dusty--2000 models  used for this  work were 
calculated as  described in  L00, and are based on the
Ames H$_2$O  and TiO line lists by  \cite{amesh2o} and \cite{amestio}.  
We  stress  that large
uncertainties persist  in these opacities for the  temperature range of
this work (see \cite{ahs00}).  The models  have been upgraded with
(i) the  replacement  of  the  JOLA  (Just Overlapping  Line
Approximation) opacities for FeH, VO  and CrH by opacity sampling line
data  from   \cite{feh}  and  R.  Freedman  (NASA-Ames,  private
communication), and (ii)  the extension of our database  of dust grain
opacities from 10  to 40 species.  These models  and their comparisons
to earlier versions  are the subject of a  separate publication (Allard
et al., in preparation).

The fitting of the synthetic to the observed spectra was done using an
automatic IDL program. First,  the resolution of the synthetic spectra
was degraded to  that  of  each individual  observed  spectrum and  the
spectra were normalized  to unit area for scaling.   The comparison was
done using  a model atmosphere  grid that covers the  range 1500~K$\leq$
$T_{\rm eff} \leq 3000$~K, $3.5\leq$$\log(g)  \leq 6.0$ and [M/H]$= -1.0, -0.5$
and 0.0,  with  a total  of  162  model  atmospheres.  For  each  observed
spectrum  we then  calculated  a  quality function  $q$,  similar to  a
$\chi^2$ value,  for the comparison  with all synthetic spectra  in the
grid.   In  order to  avoid  known  problematic  intervals in  either  the
observed  or the  synthetic spectra,  the wavelength  range 
1.5--1.7$\mu$m  was excluded  from the  comparison.  Two
particularly diagnostic wavelength
ranges, 0.7--1.4$\mu$m and 2.0--2.5$\mu$m, were given 5 times higher
weights than  the  remaining spectral  ranges.  We
selected the models that resulted in lowest 3-5 $q$ values as the most
probable  parameters range  for  each individual  star.  The  ``best''
value was  then chosen by  visual inspection. This procedure  allows a
rough estimate of  the uncertainty in the  stellar
parameters. Note
that it does not eliminate systematic errors in the stellar parameters
due to missing or incomplete opacity sources.

This  set  of models  relies  on  the assumption  that  the  dust is  in
equilibrium with the  gas phase at any depth  in the atmosphere.  This
corresponds to a limiting case, discussed also by \cite{cbah00} in the
context of brown dwarf evolution, which  could be met if the mixing of
material  (by convection,  rotation  or other  processes) prevents  the
diffusion  of the  dust.   Of course  this
situation may  not be met at  all layers of the  atmosphere and/or at
all  effective  temperatures since,  for  example,  processes such  as
convective  mixing retreat from  the upper  atmosphere as
$T_{\rm eff}$ decreases.   The current  analysis will  help
determine the  parameter range of stars over  which this approximation
is  valid.

\section{Results of Comparison of Data and Atmospheric Models}

\subsection{Low Resolution Spectra; Derived Temperatures and Diameters }

Figure 9 shows examples of our best model fits to the observed spectral
energy distributions.  Table 5 lists the best--fit model parameters,
$T_{\rm eff}$, $\log(g)$ and [m/H], for each dwarf, sorted by spectral type 
(on the Kirkpatrick scheme).  We also give diameters for the 
fourteen dwarfs with parallax measurements, derived in two ways --- one from 
the scaling factor required to match the calculated surface flux to that 
observed at the Earth, and the other from the measured luminosity and 
derived effective temperature, via the Stefan--Boltzmann law.  We return to
the diameter measurements in \S 8.

The synthetic spectra fit the energy distributions for early--L
dwarfs with $T_{\rm eff}\sim$1900~K quite well, but there are obvious
discrepancies for the hotter and cooler dwarfs. Discrepancies for
late--M dwarfs are also apparent in L00 --- the predicted water bands
are too deep and there is too much flux predicted at the peaks of the
J and H bands (in other words, the water vapor opacity profile is
too cold).  We attribute this to remaining problems in the calculated
water opacity database, which is  known to be incorrect for
higher temperature transitions (\cite{ahs00}).  There are also
discrepancies seen in the fits to the two latest spectral types in our
sample, Denis-P J 1228-1547AB  and Denis-P J 0205-1159AB.  Although
the shapes of the water bands are better matched for these objects, the
calculated fluxes are too high at the peaks of the J and H bands and
the models do not reproduce the shape of the K--band peaks. This
indicates an inaccuracy in the atmospheric structure which appears to
be too hot in the deeper layers from which the J-- and H--band flux
emerges.  These discrepancies reveal the increasing inadequacy of the
dust equilibrium assumption (or neglect of grain diffusion
mechanisms) in the model photosphere construction as $T_{\rm eff}$
decreases.  While models which incorporate gravitational settling
by diffusion are not yet available, our result demonstrates the importance
of such mechanisms in the L dwarf temperature range. Our determination 
that Denis-P J 1228-1547AB is metal--poor may only reflect problems in the Dusty
models at this $T_{\rm eff}$ --- it may in fact have solar metallicity 
(see \S 7.2).  Note that despite the improved treatment of FeH,  Figure 9 
shows that better    line data are still needed for this important opacity 
source.  Note also that investigations of the effect of using different
water opacities are presented in \cite{ahs00}. 

For this particular model grid comparison the range of $T_{\rm eff}$
allowed by the spectral energy distribution fitting is constrained to
$\pm$75--100~K, however this does not include systematic effects.  There
is further discussion of the errors in our $T_{\rm eff}$ values  in \S 8
and \S 9.
Surface gravity also affects the overall shape of the energy
distribution. The best--fit models have
$\log(g)$ of 5.5 or 6.0, except for the apparently metal--poor and
slightly lower gravity Denis-P J 1228-1547AB.  Given the remaining
uncertainties tied to the water opacity and treatment of dust, we
estimate the accuracy of our determinations of $\log(g)$ to be
0.5~dex. The structural models of \cite{cbah00} yield $\log(g)\approx$5.3
for objects in the observed temperature range with ages 1---5~Gyr.

\subsection{Medium Resolution Spectra}

Figure 10 shows the spectra we  obtained at R$\sim$3000.  Only the
J-- and K--band regions are shown; the spectra of a portion of
the H--band contain no obvious features.  Also shown are the best--fit
model synthetic spectra, based on the match to the lower resolution
data but shown at higher resolution.  We have rescaled the synthetic
spectra to match the local pseudo--continuum. 

The dominant narrow features in the J--band are the potassium
doublets.  The model spectra generally reproduce the shape and depth
of the doublet, as well as the structure to the blue of the doublet. 
The pronounced asymmetry in the K~I doublet for Denis-P J 0205-1159AB is
real, but is not understood at this time.
Most of the fine structure in the K--band is real and
is due to water absorption; the synthetic spectra reproduce the
observed structure well even at this detailed level.  Experiments with 
varying $\log(g)$ and [m/H], while keeping the effective temperature
fixed,  showed that at R$\sim$3000 the line profiles did not constrain 
gravity or metallicity significantly better than achieved with the 
R$\sim$600 data (but they did support the parameters derived from 
the other data).  For the Denis 
L dwarfs the J--band data were more diagnostic than the K--band, but 
only constrained $\log(g)$ to values between 5.0 and 5.5, and for Denis-P J
1228-1547AB showed that a metallicity of [m/H]$=0$ was slightly preferable
to the  [m/H]$=-0.5$ implied by the low resolution fit.

\section{Comparison of Data with Structural Models}

Figure 11 shows the effective temperature implied by our atmospheric model fitting
versus diameter derived by scaling the model flux to that observed at the Earth.
We also show calculated sequences for ages 0.1~Gyr and 5~Gyr from \cite{cbah00}. 
There are discrepancies between our data points and the structural model at both
the hot and cool ends of our sample.
The diameters derived here are subject to errors in the parallax measurements 
as well as in the calculated surface flux.  The variation in flux levels between 
the next--best--fit model and the best--fit model imply a typical total error
in the diameter of  10---15\%.  The error in the diameter determined from luminosity
and $T_{\rm eff}$ is dominated by the uncertainty in $T_{\rm eff}$ and is
typically 15\% if the error in $T_{\rm eff}$ is $\sim$100~K.  These are the errors
shown by the error bars in Figure 11.

However as discussed above there are large uncertainties remaining in the 
opacity treatment for the  temperature range of this work (see \cite{ahs00}), 
and in the treatment of dust --- the assumption here that the dust is  in
equilibrium with the  gas phase at any depth  in the atmosphere
corresponds to a limiting case, discussed also by \cite{cbah00}.  It is possible
that systematic errors exist in our derivations of effective
temperature and diameter, despite our exclusion of the most suspect wavelength
regions in the model fitting.  

The theoretical determinations of radii and luminosity for low--mass stars and 
brown dwarfs are thought to be reasonably robust and reasonably
independent of the surface atmosphere used for the structural model.
This is demonstrated by the fact that the radii calculated by \cite{bur97}
for brown dwarfs of mass 0.03---0.07~M$_{\odot}$ with 1300 $\leq T_{\rm
eff}$~K$ \leq$ 2600, and age 0.3---3.0~Gyr, range (only) from 5.8e7~m to 
7.6e7~m, in close agreement with
those derived by \cite{cbah00} of 6.0---7.6e7~m.  Figure 12 plots $T_{\rm eff}$
versus log$_{10}$L/L$_{\odot}$ for objects ranging in mass from 
0.03~M$_{\odot}$ to 0.10~M$_{\odot}$ and aged 0.1~Gyr to 10~Gyr.  These
are a reasonable to generous range in parameters for our observed range in 
luminosity.  The sequences shown are from \cite{cbah00}, but one mass
sequence from \cite{bur97} is also shown to demonstrate the good agreement
between evolutionary models.  We have used this figure to derive effective
temperatures for the targets in the sample with known distance (and hence
luminosity) and these values are listed in the last column of Table 5.
We assume all objects are single except for Denis-P J 1228-1547AB and 
Denis-P J 0205-1159AB, for which we assume two identical components.
Note that the older kinematic population class for LHS 429 would suggest
that the effective temperature lies at the high end of the tabulated
range for this object.

There is an indication that our effective temperatures derived from fitting
the spectral energy distributions with the synthetic spectra may be 
underestimated for the hotter stars in our sample, with spectral type
around M7, and overestimated for the cooler mid-- to late--L dwarfs.  A
hotter scale for the late--M dwarfs would agree better with that derived in
\cite{l00} (i.e. $T_{\rm eff} \sim 2600 $~K for type dM6). 
The agreement between these two approaches --- the spectral
synthesis and luminosity comparison ---  is $\sim$200~K, except for the
coolest object in the sample,   Denis-P J 0205-1159AB, where the
discrepancy is $\sim$400~K.  Better determinations of effective temperature
for objects this late will have to await models with improved treatment of 
dust.

\section{Discussion of the L--Dwarf Temperature Scale and Conclusions}

Our derived temperatures agree  with other derivations of
$T_{\rm eff}$ for early--L type dwarfs within the likely errors.
\cite{pav00} use both the NextGenDusty  and \cite{tsu00}
atmospheric structures with their
own spectral synthesis code to fit red spectra; they have to include
an unknown additional opacity source, in the form of a power--law, to
match the overall spectral shape because they do not include dust in
solving for chemical equilibrium,  or account for the opacity
of known dust species (note that this is inconsistent with their adopted 
atmospheric structure).
They derive $T_{\rm eff}$ for four dwarfs in common with us: for
BRI 0021-0214 and Kelu-1 their values are higher than our spectral 
synthesis values by 100~K; for Denis-P J 1228-1547AB and Denis-P J 0205-1159AB 
their values are 200~K and 700~K lower.  Their value of 1200~K for Denis-P J 
0205-1159AB does not seem likely, as we would then expect to see methane 
features  in the H-- and K--band regions, which are not seen. 

\cite{bas00} use Ames--Cond NextGen model fits to rubidium and
cesium lines and derive effective temperatures   within 100~K of our 
spectral synthesis values for Kelu-1, Denis-P J 1058-1548 and Denis-P J 1228-1547AB; 
they derive a value 150~K lower than ours for the coolest object in common,
Denis-P J 0205-1159AB, and a value 300~K higher for the M9.5 dwarf LP 944-20.
The Ames-Cond models include the formation of dust but neglect its
opacity (i.e. assume the dust has settled below the photosphere),
which corresponds to a limiting case which complements that
assumed here of the dust in equilibrium with the gas phase.
It is expected that dust clouds will sink lower in the atmosphere with
decreasing effective temperature (in the absence of replenishment of 
material in the upper photosphere), hence the Cond
models are likely to be more correct than the Dusty models
for the latest L--dwarfs. The Basri et al. lower temperature
of 1750~K for Denis-P J 0205-1159AB is likely to be more accurate than that 
derived here, but the opposite is true at hotter temperatures and so 
our lower $T_{\rm eff}$ for LP 944-20 is likely to be more accurate than 
their value.  Given the discrepancies seen in the
model fits to the observed spectral energy distributions for the coolest 
objects in our sample, it would appear that gravitational settling of 
dust becomes important at $T_{\rm eff}\leq1800$~K.

Based on species condensation arguments, \cite{k00} argue
for an effective temperature around 2000~K for L0 spectral types,
which is in agreement with this work.  They also advocate (based on
luminosity) $T_{\rm eff}\sim$1300~K for the L8 dwarf (by the Kirkpatrick
scheme) GL 584C. However the discovery of three dwarfs optically
redder than L8 and showing both CO and CH$_4$ in their spectra
(\cite{l00}) suggests that these new later--type objects have 
$T_{\rm eff}\sim$1300~K and by implication L8 spectral types have
$T_{\rm eff}>$1300~K.  A hotter temperature scale also is supported by the
preliminary model fits to L--band methane features seen in mid--L
dwarfs reported by \cite{noll00}. However all of these results can be estimates
only as they either rely on the correlation between the appearance or 
disappearance of various species, which depends on the temperature of the 
line--forming region and not on the effective temperature, or they rely
on what are acknowledged to be very preliminary atmospheric models
in this temperature range.

In conclusion, the best current estimates are that the M dwarfs cover 
a range in $T_{\rm eff}$ of 3700---2100~K, the L dwarfs 2000---$\sim$1500~K 
and the known methane (T) dwarfs $\sim$1300---800~K.  A definitive temperature 
scale for the L dwarfs will not be possible
until model atmospheres which include gravitational settling
are available; such work is in progress by the modellers of this group
and others.  We note that any object with a spectral
type L5 or later  must be substellar based on evolutionary
models, and late--M to early--L dwarfs may be substellar depending on
their age.  The discovery of binary L--dwarf systems offers a chance
to further improve our knowledge of the fundamental parameters for
this recently discovered low--mass population of our Galaxy.

\acknowledgments We are very grateful to the staff at UKIRT, and at the
University of Hawaii 2.2~m (88~inch) telescope, for their
assistance in obtaining the data presented in this paper.  Some of
these data were obtained through the UKIRT Service Programme.  UKIRT
is operated by the Joint Astronomy Centre on behalf of the
U.K. Particle Physics and Astronomy Research Council.  We are
grateful to Richard Freedman (NASA-Ames) who generously provided VO and
CrH line lists for use in the current
models. FA  acknowledges support from CNRS.  PHH
acknowledges partial support from the P\^ole Scientifique de
Mod\'elisation Num\'erique at ENS-Lyon.  
This work was supported in part by NSF grant
AST-9720704, NASA ATP grant NAG 5-8425 and LTSA grant NAG 5-3619, as
well as NASA/JPL grant 961582 to the University of Georgia.
Some of
the calculations presented in this paper were performed on the IBM SP
and the SGI Origin 2000 of the UGA UCNS, on the IBM SP of the San
Diego Supercomputer Center (SDSC, with support from the National
Science Foundation), on the IBM SP and Cray T3E of the NERSC with support 
from the DoE, on the IBM SP2 of the French Centre National Universitaire
Sud de Calcul (CNUSC).  We thank all these institutions for a generous
allocation of computer time.

\newpage 

\begin{figure}

\plotfiddle{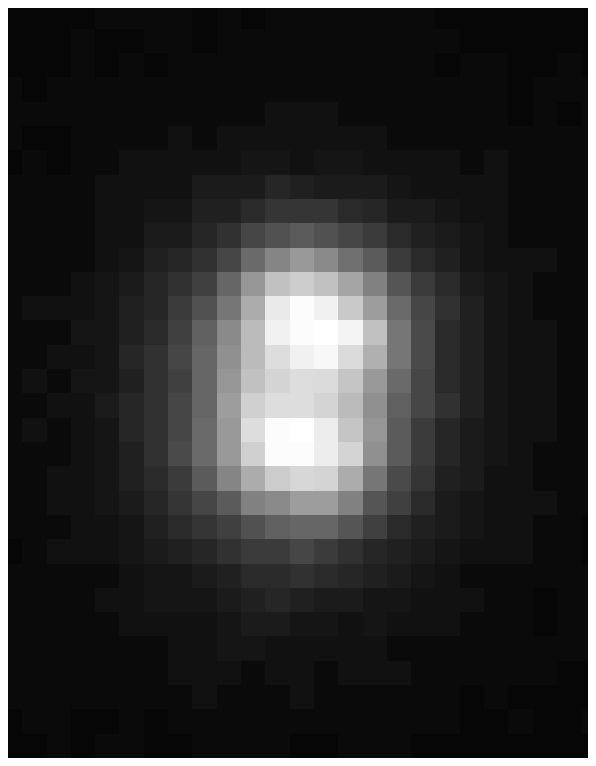}{10truecm}{-90}{100}{100}{-400}{450}
\caption{K band image of Denis-P J 0205-1159.  The pair are separated by
0$\farcs$35; North is up, East to the left.
\label{fig1}}
\end{figure}

\newpage

\begin{figure}
\epsscale{.8}
\plotone{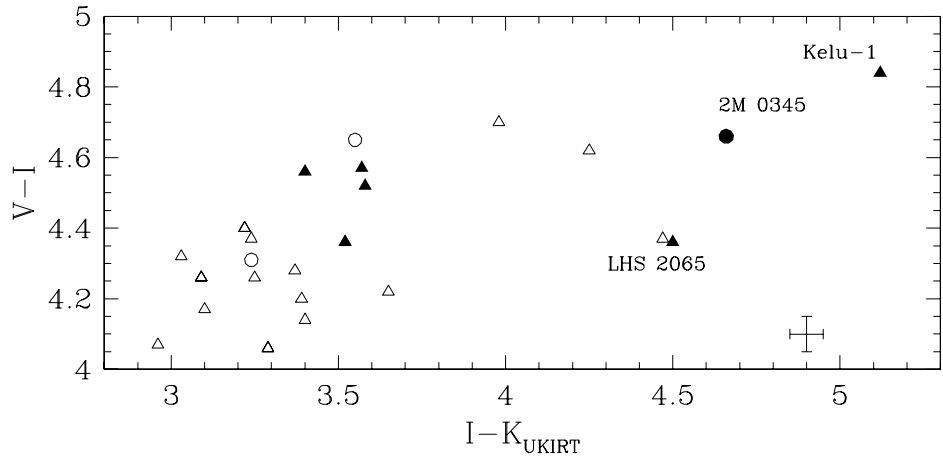}
\caption{V$-$I:I$-$K diagram; filled symbols are
this work, open symbols are dwarfs from \cite{l98,l92} 
(on the UKIRT JHK system).
Symbol shapes represent kinematic populations: 
triangles --- disk, circles --- unknown.   Typical 
uncertainties are shown at lower right. 
\label{fig2}}
\end{figure}

\newpage

\begin{figure}
\epsscale{.8}
\plotone{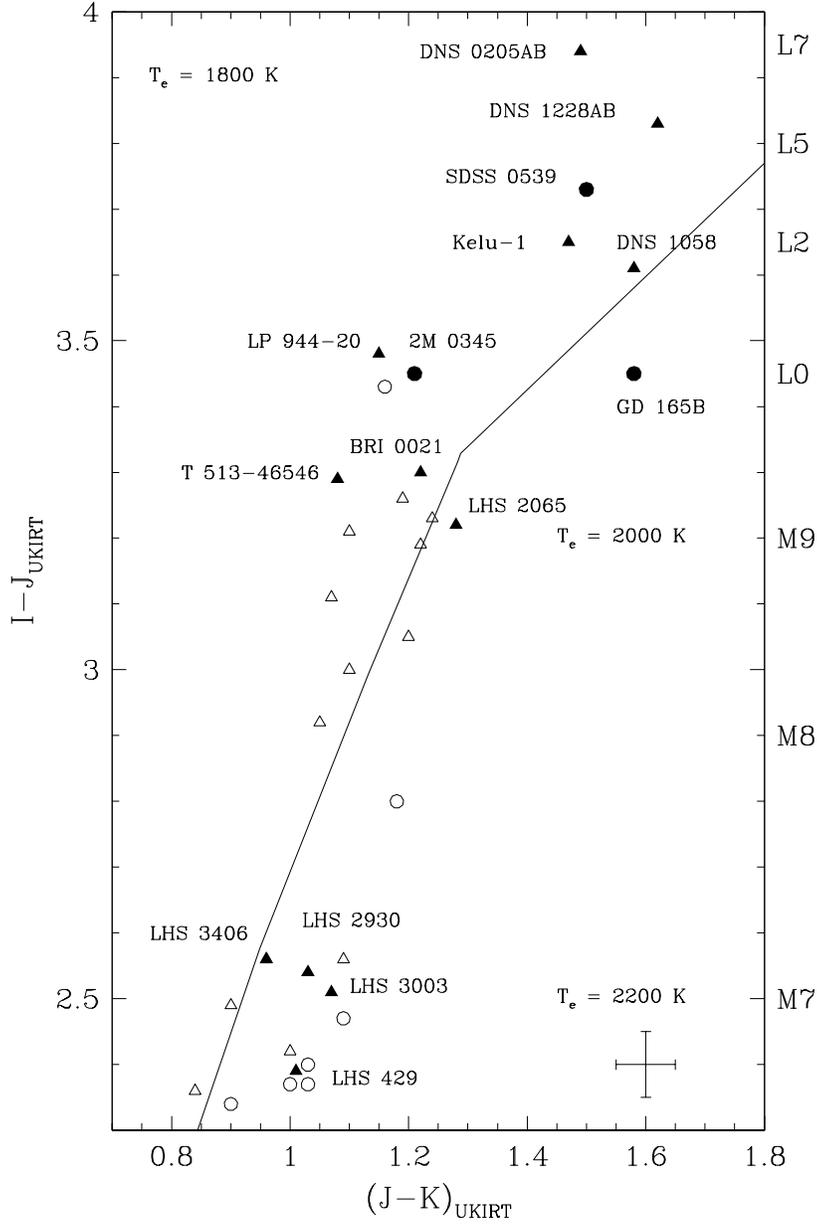}
\caption{I$-$J:J$-$K diagram using the same  symbols 
as in Figure 2. Spectral types based on I$-$J 
are shown for guidance only --- for the
L dwarfs there is significant variation of color with type.  
The solid line is the Dusty sequence for [m/H]$=$0 and
$\log(g)=$5.0 as a function of effective temperature.
\label{fig3}}
\end{figure}

\newpage

\begin{figure}
\epsscale{.8}
\plotone{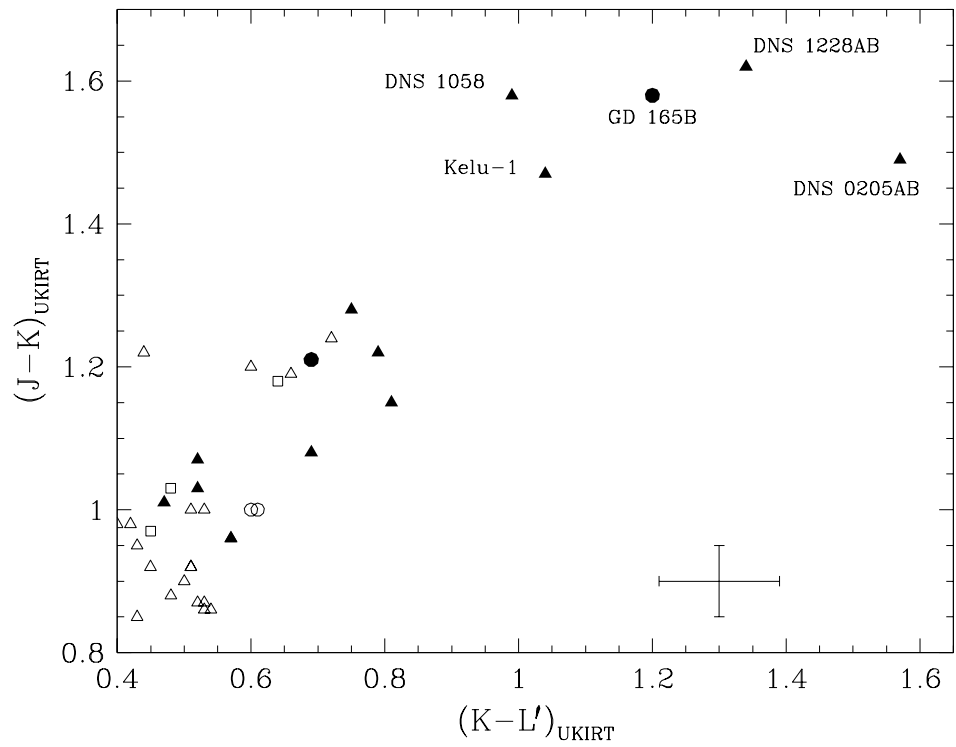}
\caption{J$-$K:K$-$L$^{\prime}$ diagram, using the same symbols as in Figure 2.
\label{fig4}}
\end{figure}

\newpage

\begin{figure}
\epsscale{.8}
\plotone{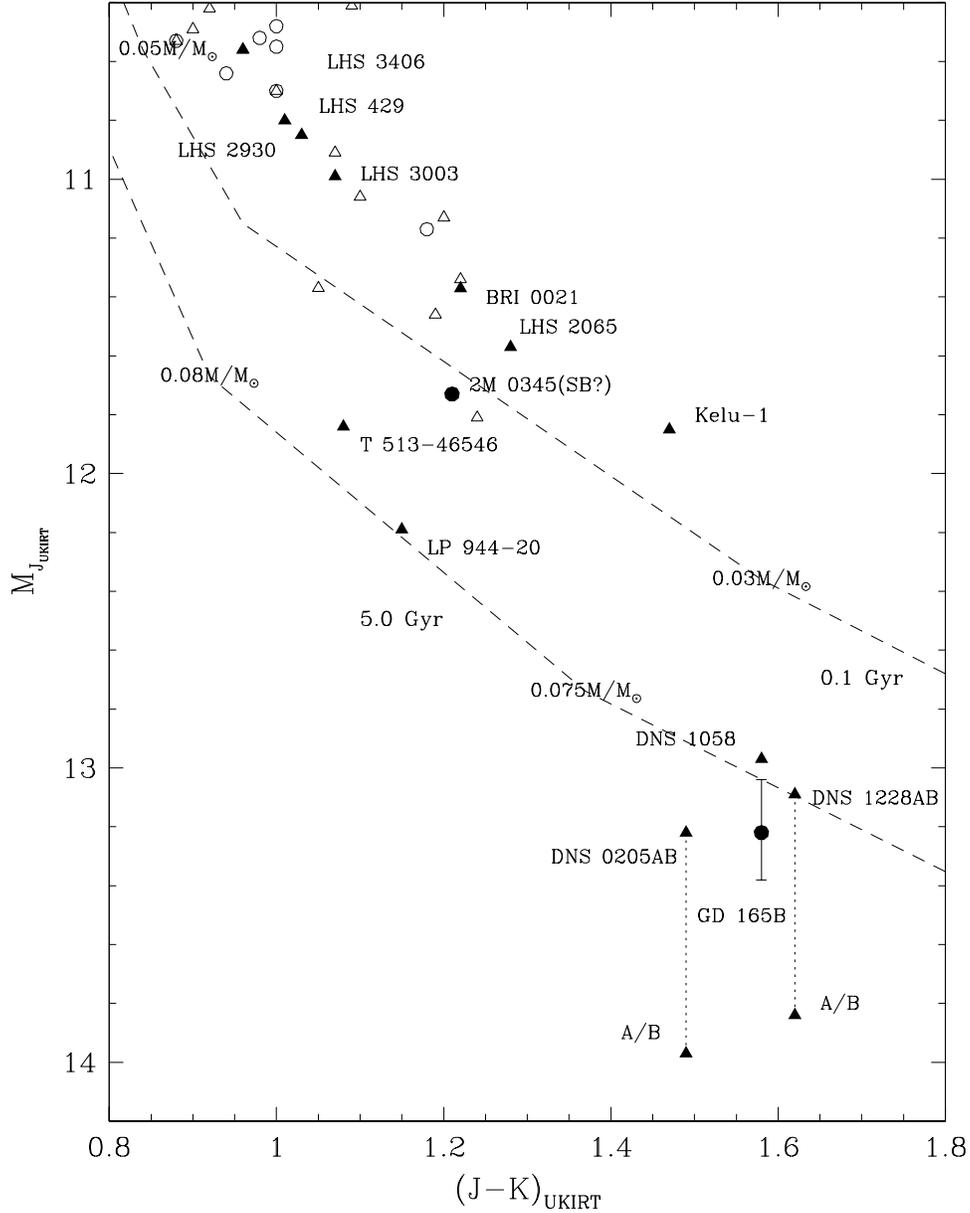}
\caption{M$_J$:J$-$K with symbols  as in Figure 2. The location of the
identical components of Denis-P J 0205-1159 and Denis-P J 1228-1547, if they 
were resolved, are indicated by the dashed line and lower triangles.  The 
isochrones for 0.1~Gyr and 5.0~Gyr are from \cite{cbah00}.
\label{fig5}}
\end{figure}

\newpage
\begin{figure}
\plotfiddle{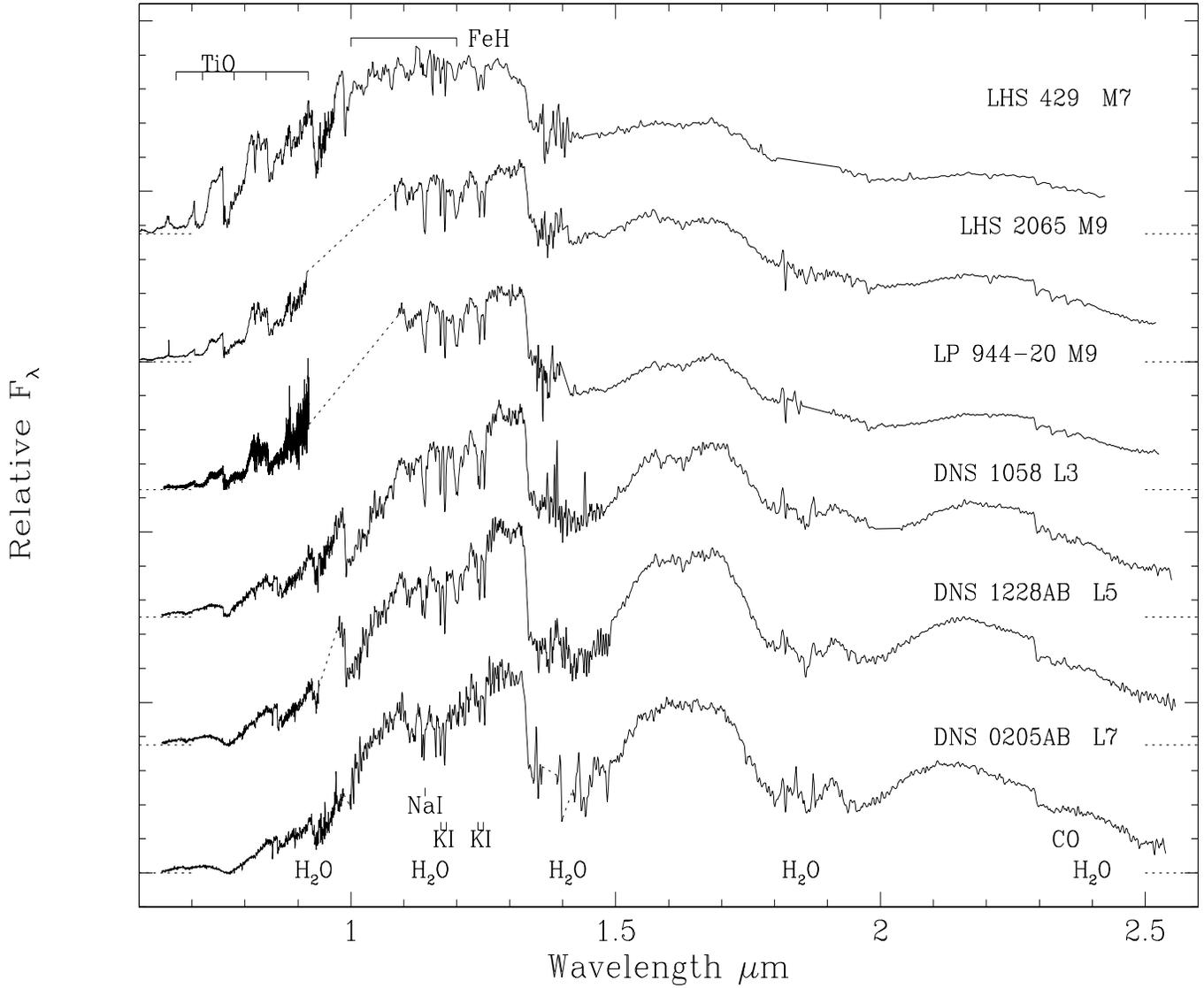}{10truecm}{-90}{70}{80}{-270}{450}
\caption{Spectral sequence for late--M to L dwarfs where the spectra have been 
normalised to the flux at 1.2$\mu$m and offset.  
Horizontal dotted lines indicate zero flux levels.  
Denis-P J 0205-1159 is classified L5 in the Mart\'{\i}n scheme. 
\label{fig6}}
\end{figure}

\newpage

\begin{figure}[t!]
\plotfiddle{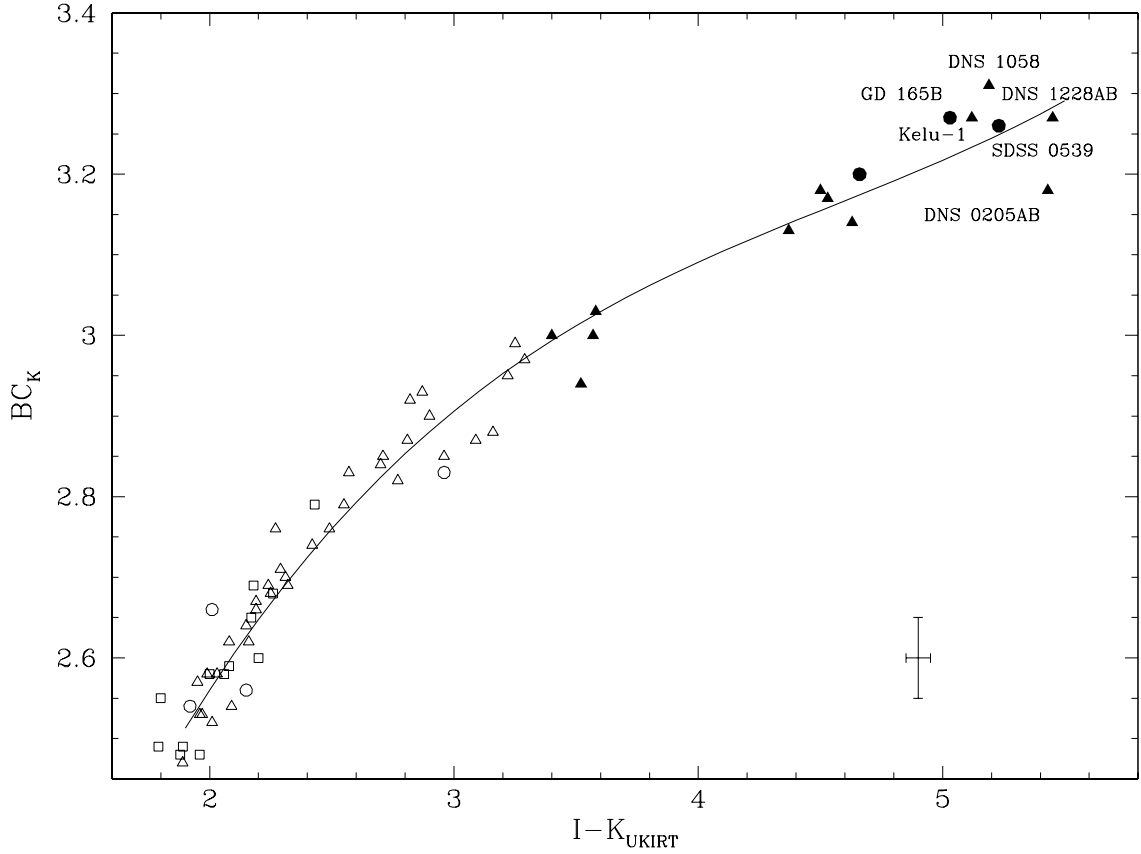}{10truecm}{-90}{60}{60}{-220}{350}
\caption{Bolometric correction at K versus I$-$K.  Filled symbols are
this work, open are from Leggett et al. 1996, 2000a. Symbol shapes
represent kinematic 
populations: squares --- halo, triangles --- disk, circles --- unknown.
The solid line is the empirical fit to the disk objects.
\label{fig7}}
\end{figure}

\newpage

\begin{figure}[t!]
\plotfiddle{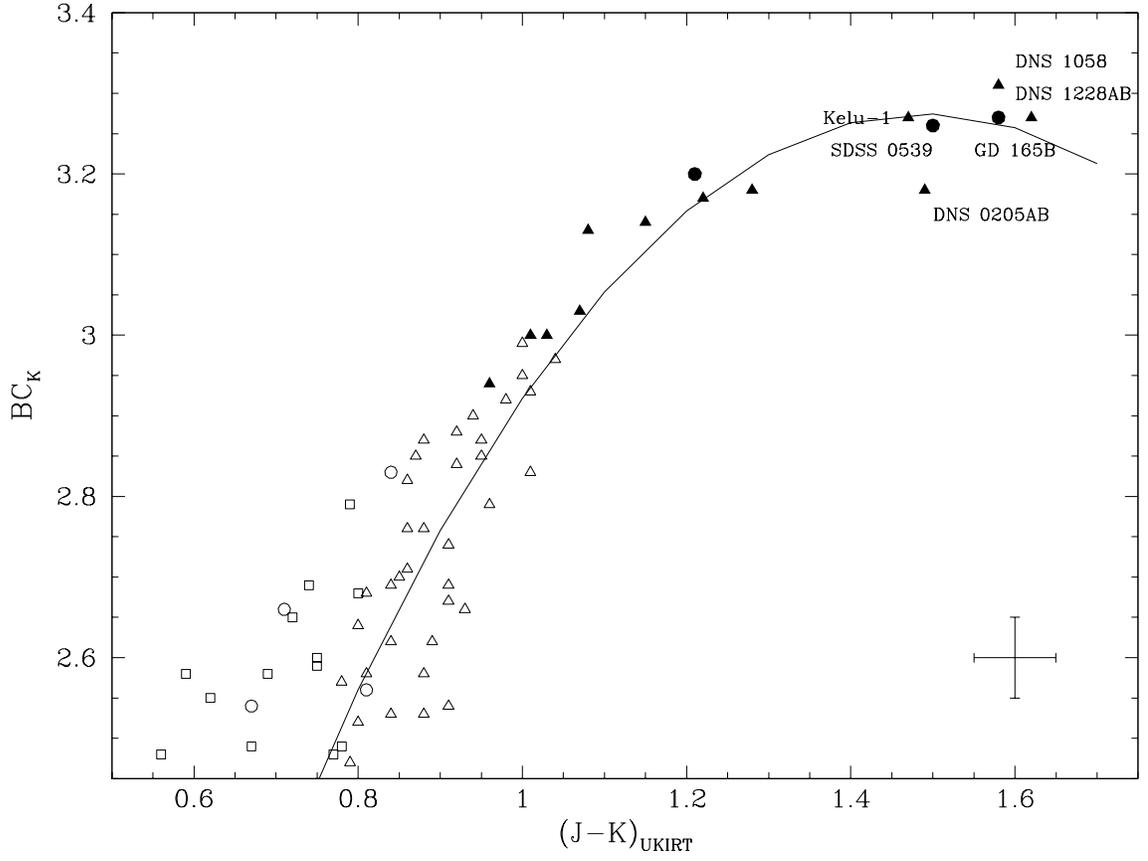}{10truecm}{-90}{60}{60}{-220}{350}
\caption{Bolometric correction at K versus J$-$K. Symbols are
as in Figure 7.  The solid line is the empirical fit to the disk objects
(circles and triangles).
\label{fig8}}
\end{figure}

\newpage
\begin{figure}
\plotfiddle{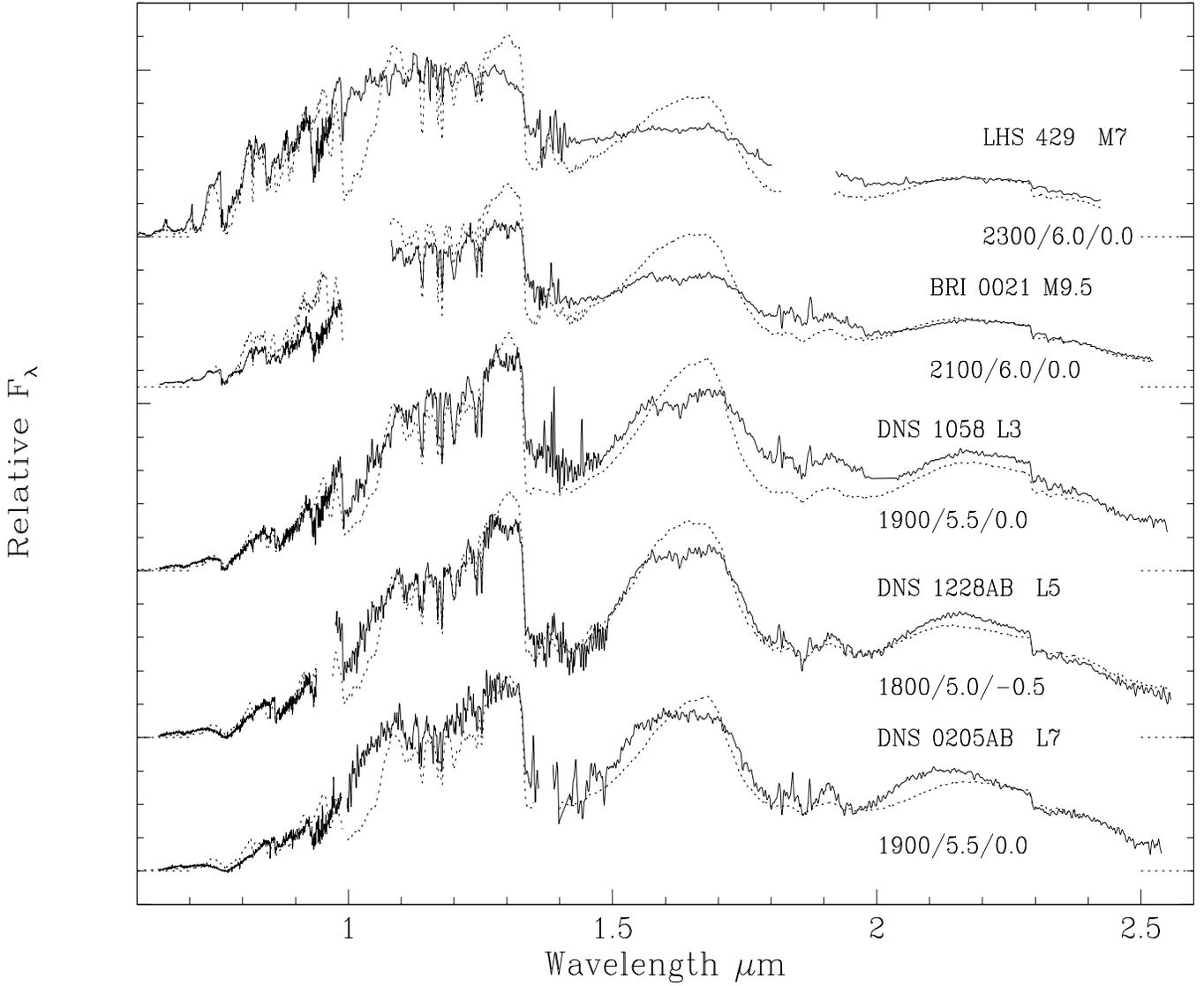}{10truecm}{-90}{70}{80}{-270}{450}
\caption{Examples of best synthetic spectra fits to late--M and L dwarfs,
where the models are shown as dotted lines and 
$T_{\rm eff}$/$\log(g)$/[m/H] are given.
The spectra have been normalised to the flux at 1.2$\mu$m and offset. 
Horizontal dotted lines indicate zero flux levels.  
Denis-P J 0205-1159 is classified L5 in the Mart\'{\i}n scheme. 
\label{fig9}}
\end{figure}

\newpage
\begin{figure}
\plotfiddle{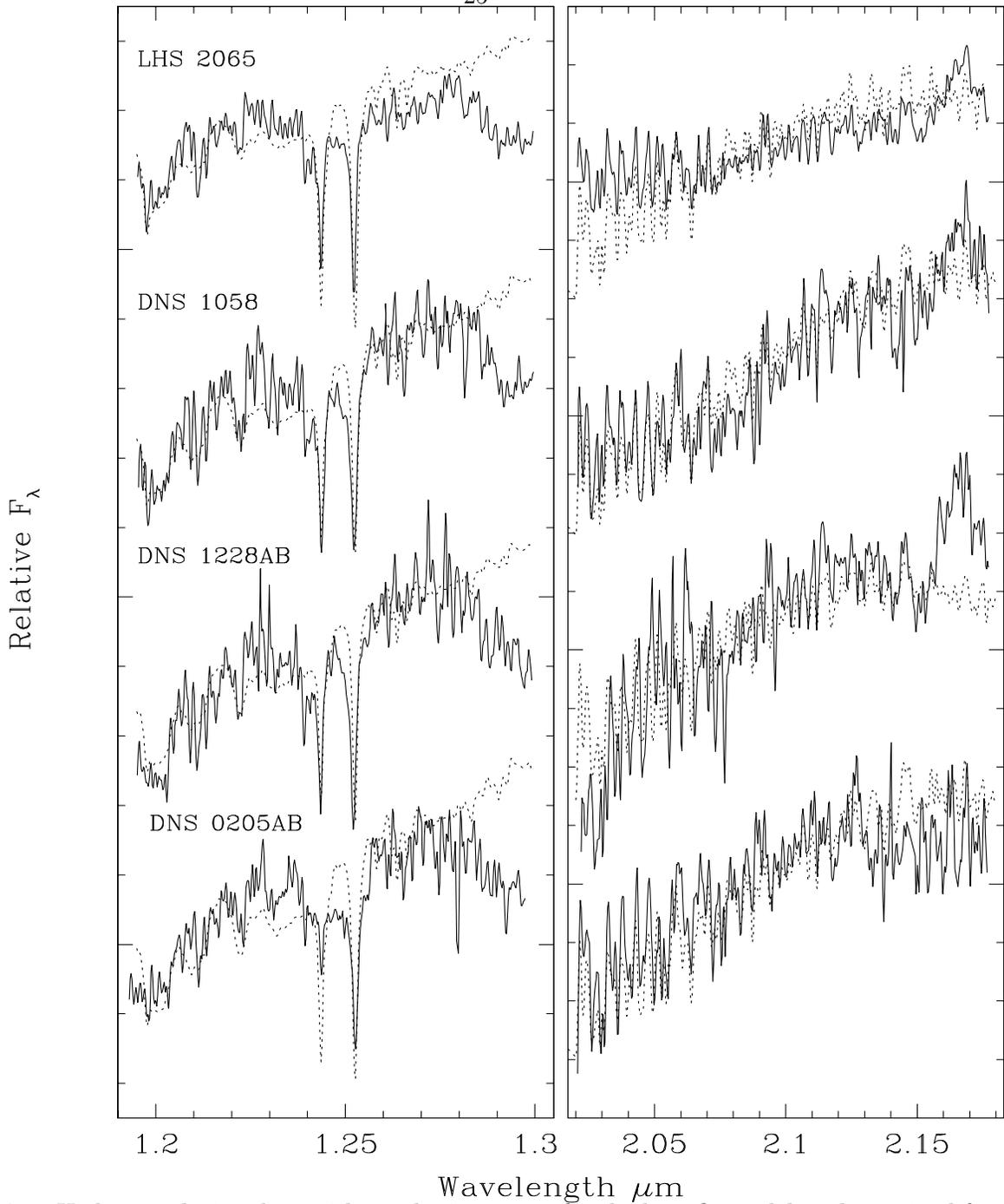}{14truecm}{0}{80}{75}{-220}{-50}
\caption{Higher resolution data with synthetic spectra for the best fit model
as determined from the low resolution spectral energy 
distributions.  The spectra have been normalised and offset for clarity.
Much of the structure in the K--band is real and due to water.
\label{fig10}}
\end{figure}

\newpage
\begin{figure}
\plotfiddle{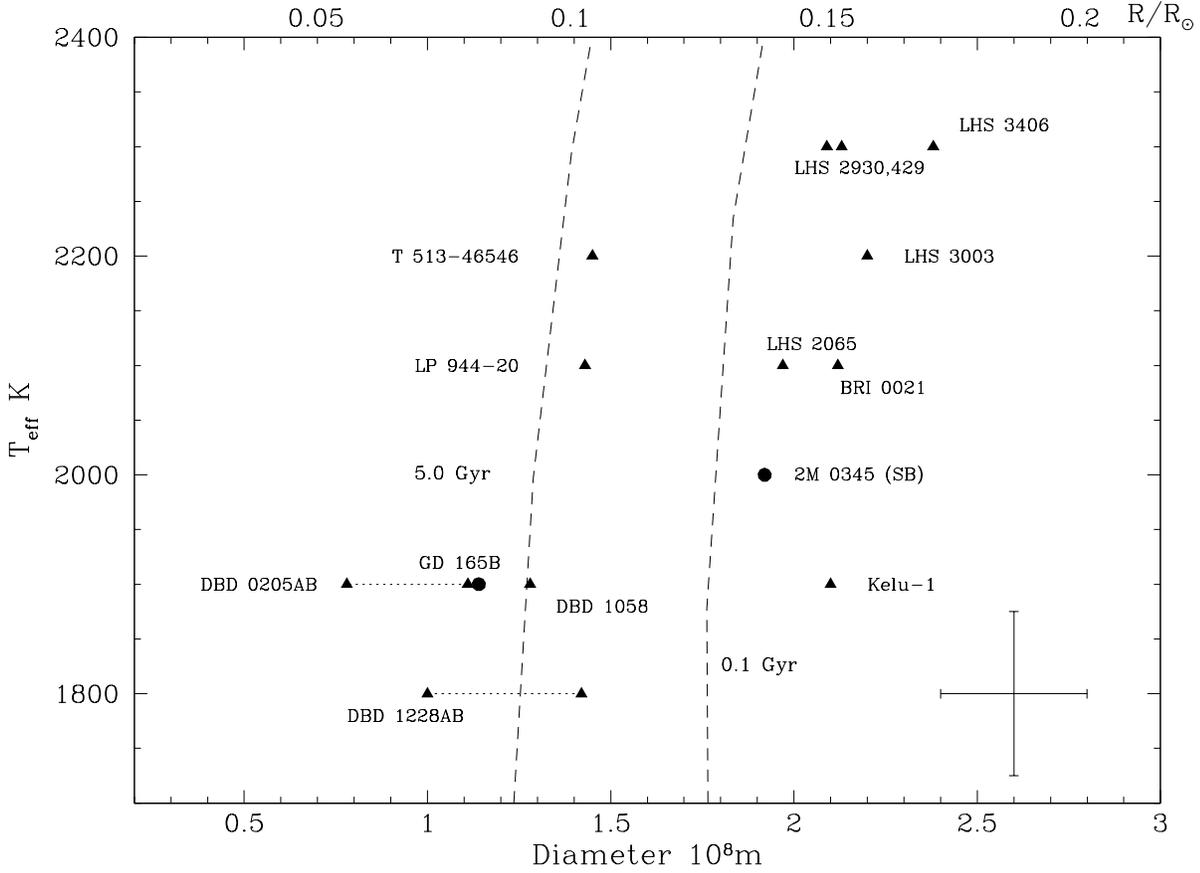}{10truecm}{-90}{60}{60}{-250}{350}
\caption{Diameter (derived by scaling) as a function of  $T_{\rm eff}$,
from the model atmosphere comparison.
Dashed lines are structural model calculations from \cite{cbah00} for
0.1~Gyr and 
5.0~Gyr.  The location of the identical components of Denis-P J 0205-1159 and 
Denis-P J 1228-1547, if they were resolved, are indicated by the dashed lines to
smaller diameters.  
\label{fig11}}
\end{figure}

\newpage
\begin{figure}
\plotfiddle{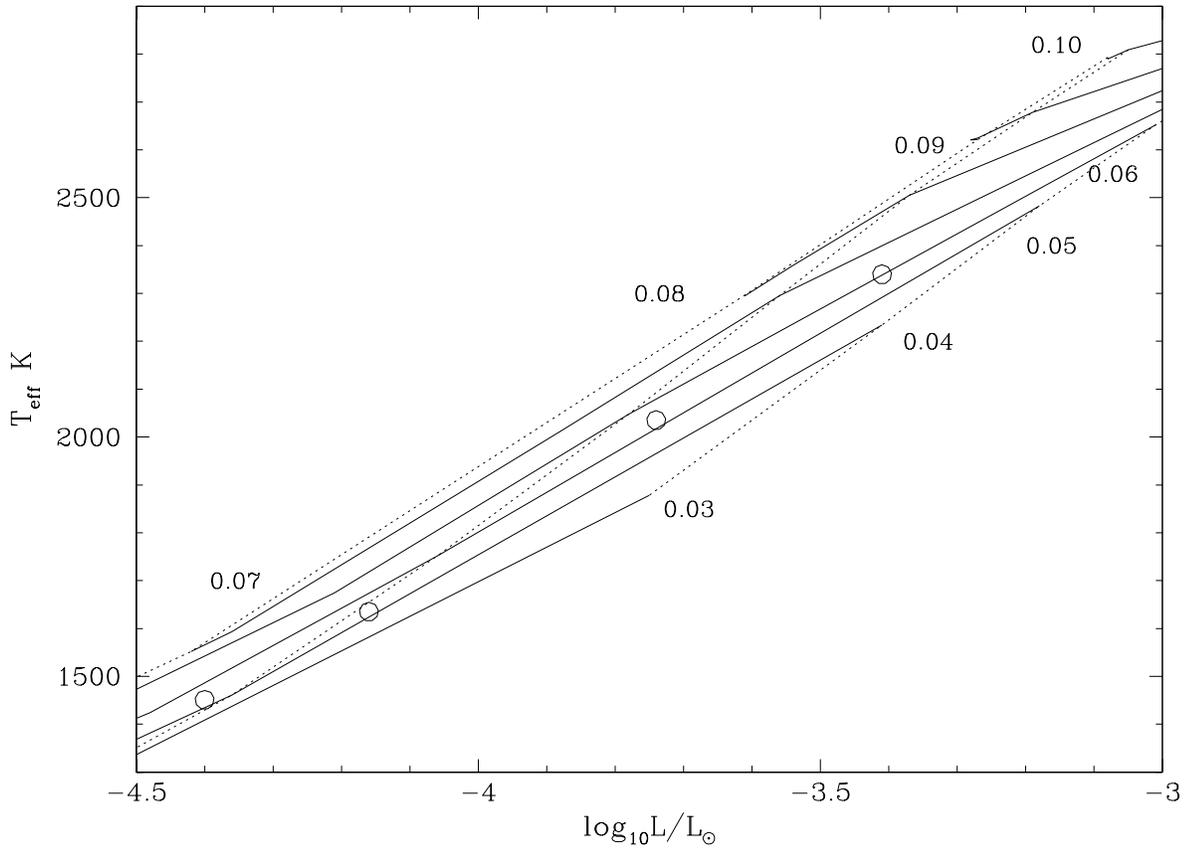}{10truecm}{-90}{60}{60}{-250}{350}
\caption{Luminosity as a function of  $T_{\rm eff}$ for constant mass,
from the structural models of \cite{cbah00}.  Masses are indicated in units of
M$_{\odot}$ and evolution proceeds right to left; dotted lines indicate
0.1, 1.0 and 10.0 Gyr.  The circles indicate the 0.04~M$_{\odot}$ sequence
from \cite{bur97} for comparison.
\label{fig12}}
\end{figure}

\clearpage

\begin{deluxetable}{rlclclr}
\tablenum{1}
\tablecaption{Dwarfs in the Sample}
\tablehead{
\colhead{LHS} & \colhead{Other} & \colhead{RA/Dec} & 
\colhead{Spectral\tablenotemark{a}} &\colhead{Kinematic} & 
\multicolumn{2}{r}{Instrumentation\tablenotemark{b}} 
\nl
\colhead{Number}  & \colhead{Name} & \colhead{1950} &
\colhead{Type} & \colhead{Population} & \colhead{Infrared}  
& \colhead{Optical } \nl  
}
\tablecolumns{7}
\startdata 
\nodata  & BRI 0021-0214 & 0021$-$02 &  M9.5 & Y/O   &  IR1 & 1 \nl
\nodata & Denis-P J 0205-1159AB & 0203$-$12 & L7 &  YD & IR1,IR0
& 1 \nl
\nodata  & LP 944-20 & 0337$-$35 & M9  &  YD   &   IR1& 2 \nl
\nodata  & 2MASP J0345432$+$254023 & 0342$+$25 & L0 & \nodata & IR1& 3\nl
\nodata  & SDSS 0539 & 0539$-$00 & L5 & \nodata & 4 & 5 \nl
2065 & LP 666-9 & 0851$-$03 & M9 &  YD  & IR1,IR0 & 6 \nl
\nodata   &  Denis-P J 1058-1548   &      1058$-$15  &  L3   &   YD  &
IR1,IR0 & 1\nl
\nodata   &  Denis-P J 1228-1547AB  & 1228$-$15 & L5 &  YD  & IR1,IR0
& 1\nl
\nodata  &  Kelu-1 & 1305$-$25 & L2 & YD  & IR3& 7 \nl
\nodata  &  GD 165B & 1422$+$09 & L4 &  \nodata & IR3& 6 \nl
2930  &  LP 98-79   &      1429$+$59  &  M6.5   &   Y/O  &   IR4& 8 \nl
3003  &  LP 914-54    &     1453$-$27  &   M7    &     Y/O  &  IR4 & 9\nl
\nodata   &  TVLM 513-46546   &    1458$+$23  &   M8.5   &  Y/O  &  IR3& 9\nl
 429  &  Gl 644C   &      1652$-$08  &  M7     &   OD   &   IR4 & 9\nl
3406  & LP 229-30 & 1841$+$40 & M5.5 &   Y/O & IR4 &  10\nl

\tablenotetext{a}{Spectral types quoted on the Kirkpatrick scheme, note that
Denis-P J 0205-1159AB is classified L5 on the Mart\'{\i}n scheme, see text}
\tablenotetext{b}{Published spectra taken from: \\
1 Tinney et al. 1998 resolution 7A \\
2 Tinney \& Reid 1998 resolution 0.4A \\
3  Kirkpatrick, Beichman \& Skrutskie  1997 resolution 12A \\
4 Leggett et al. 2000b resolution 25---50A\\
5 Fan et al. 2000 resolution 14A \\ 
6 Kirkpatrick et al. 1995 resolution 18A\\
7 Ruiz et al. 1997 resolution 17A\\
8 Kirkpatrick 1992 resolution 18A\\
9  Kirkpatrick et al. 1995 resolution 12A\\
10 Reid, Hawley \& Gizis 1995 resolution 8A\\
}
\enddata

\end{deluxetable}

\newpage
\begin{deluxetable}{cllcrcrrrr}
\tablenum{2}
\tablewidth{0pt}
\tablecaption{Instrumentation for New Spectroscopic Data}
\tablehead{
\colhead{Configuration} & \colhead{Instrument} & \colhead{Telescope} & 
\colhead{Date} &\colhead{Grating} & \colhead{Slit} &
\multicolumn{4}{c}{Resolution A} \nl
\colhead{Name}  & \colhead{} & \colhead{} & \colhead{YYYYMM} & 
\colhead{lines/mm} & \colhead{Width \arcsec} &  \colhead{Z}& 
\colhead{J} & \colhead{H} & \colhead{K}\nl  
}
\tablecolumns{10}
\startdata 
IR0  &  CGS4  &  UKIRT & 199801 & 150 & 1.2 & \nodata &   4
&  6  &  6 \nl
IR1  &   CGS4    &     UKIRT    &  199904 &  40       &  0.6 
&  \nodata &  17  &  35  &  35\nl
& & & 199801 & & & &  & \nl
& & & 199711 & & &  & & \nl
IR3  &  CGS4     &    UKIRT     &   199704  &     75   &  1.2 
& \nodata   &     16  &  30  &  30\nl
IR4  &  KSPEC    &    UH 88"   &   199407  & 60  &  1.2   
&    20   &  25  &  30  &  40\nl
\nl
\enddata
\end{deluxetable}

\newpage

\begin{deluxetable}{lrrrr}
\tablenum{3}
\tablewidth{400pt}
\tablecaption{New Photometry}
\tablehead{
\colhead{Name} & \colhead{J (error)} & \colhead{H (error)} & 
\colhead{K (error)} &\colhead{L$^{\prime}$ (error)} \nl
\colhead{} & \multicolumn{3}{c}{UKIRT--UFTI(MKO--NIR)} & \colhead{UKIRT}
}
\tablecolumns{5}
\startdata 
Denis-P J 0205-1159AB & 14.43 (0.04) & 13.61 (0.03)  & 12.99 (0.03) &
11.44 (0.10) 
\nl
2MASP J0345432$+$254023  & \nodata & \nodata & \nodata   & 12.01 (0.10) \nl
Denis-P J 1058-1548   & \nodata  & \nodata  & \nodata  & 11.62 (0.10) \nl
Denis-P J 1228-1547AB  & \nodata  & \nodata  & \nodata   & 11.42 (0.10) \nl
\enddata

\end{deluxetable}

\newpage

\hoffset=-3.0truein
\begin{deluxetable}{crrrrrrrrrrrr}
\footnotesize
\tablenum{4}
\tablewidth{550pt}
\tablecaption{Colors (on the Cousins and UKIRT--IRCAM3 systems) and Fluxes}
\tablehead{
\colhead{Name} & \colhead{$M-m$\tablenotemark{a}} & 
\colhead{V$-$I} & \colhead{I} 
& \colhead{I$-$J} & \colhead{J$-$H} 
& \colhead{H$-$K} & \colhead{K} 
& \colhead{K$-$L$^{\prime}$} & \colhead{Flux\tablenotemark{b}} & 
\colhead{m$_{bol}$\tablenotemark{c}} & \colhead{BC$_{K}$\tablenotemark{d}} & 
\colhead{log $_{10}$\tablenotemark{e}} \nl
\colhead{}  & \colhead{}    & \colhead{} & \colhead{}
& \colhead{}    & \colhead{}   & \colhead{}   & \colhead{} 
& \colhead{} &  \colhead{W/m$^2$} & \colhead{}
& \colhead{} &  \colhead{L/L$_{\odot}$} \nl
}
\startdata 
BRI 0021-0214 & $-0.42$\tablenotemark{1} & \nodata & 15.10 & 3.30 & 0.76 &
0.46 & 10.57 & 0.79 &   8.17e$-$14  &  13.74    &  3.17    & $-$3.43 \nl
Denis-P J 0205$-$1159AB & $-1.28$\tablenotemark{2} & \nodata & 18.44 & 3.94 &
0.98 & 0.51 & 13.01 & 1.57 & 8.60e$-$15 & 16.19 & 3.18 & $-4.06$  \nl
LP 944-20  & 1.51\tablenotemark{1} & \nodata & 14.16  & 3.48 &  0.70 & 0.45
& 9.53 & 0.81 & 2.20e$-$13 & 12.67 & 3.14 & $-$3.77 \nl
2MASP J0345432$+$254023 & $-2.18$\tablenotemark{2} & 4.66 & 17.36 & 3.45 & 0.77 &
0.44 & 12.70 & 0.69 & 1.12e$-$14 & 15.90 & 3.20 & $-3.59$ \nl
SDSS 0539 & \nodata & \nodata & 17.67 & 3.73 & 0.97 & 0.53 & 12.44 & \nodata &
1.34e$-$14 & 15.70 & 3.26 & \nodata \nl
LHS 2065 &  0.35\tablenotemark{3} &  4.36  & 14.44  & 3.22  & 0.80  & 0.48  &
9.94 & 0.75 & 1.45e$-$13 & 13.12 & 3.18 & $-$3.49 \nl
Denis-P J 1058$-$1548 & $-1.22$\tablenotemark{2} & \nodata & 17.80 &  3.61 & 0.99
& 0.59 & 12.61 & 0.99 & 1.10e$-$14 & 15.92  & 3.31 & $-$3.98 \nl
Denis-P J 1228$-$1547AB & $-1.29$\tablenotemark{2} & \nodata & 18.21  & 3.83 &
1.05 & 0.57 & 12.76 & 1.34 & 9.94e$-$15 & 16.03   &  3.27 & $-$4.00   \nl
Kelu-1 & $-1.44$\tablenotemark{2} & 4.84 & 16.94 & 3.65 & 0.92 & 0.55 &
11.82 & 1.04 & 2.37e$-$14 & 15.09 & 3.27 & $-$3.56 \nl
GD 165B&  $-2.49$\tablenotemark{3} & \nodata & 19.16 & 3.45 & 1.03 & 0.55 &
14.13 & 1.20 & 2.82e$-$15 & 17.40 & 3.27 & $-$4.06 \nl
LHS 2930 & 0.08\tablenotemark{3} & 4.57 & 13.31 & 2.54 & 0.65 & 0.38 & 9.74 &
0.52 & 2.05e$-$13 & 12.74 & 3.00  &$-$3.23  \nl
LHS 3003 & 0.97\tablenotemark{3} & 4.52 & 12.53 & 2.51 & 0.65 & 0.42 & 8.95
& 0.52 & 4.14e$-$13 & 11.98 & 3.03 & $-$3.28\nl
TVLM 513-46546 & 0.04\tablenotemark{1} & \nodata & 15.09 & 3.29 & 0.68 & 0.40
& 10.72 & 0.69 & 7.43e$-$14 & 13.85 & 3.13 & $-$3.65 \nl
LHS 429 &  0.95\tablenotemark{3} &  4.56 & 12.24 & 2.39 & 0.63 & 0.38 & 8.84 &
0.47 & 4.69e$-$13 & 11.84 & 3.00 & $-$3.22 \nl
LHS 3406  &  $-0.75$\tablenotemark{3} & 4.36  & 13.87 & 2.56 & 0.61 & 0.35 &
10.35 & 0.57  & 1.24e$-$13 & 13.29 & 2.94 & $-$3.12  \nl
\nl
\tablenotetext{a}{Distance moduli from: \nl
1: Tinney 1996 \nl
2: Dahn et al. 2000 \nl
3: Yale General Catalogue (van Altena et al. 1995) }
\tablenotetext{b}{Flux at the Earth}
\tablenotetext{c}{Apparent bolometric magnitude; adopting L$_{\odot} = 3.86e26$ W and 
M$_{bol\odot} = 4.75$ then:
$ m_{bol} = -2.5\times log_{10}(flux) - 18.978 $}
\tablenotetext{d}{BC$_K = m_{bol} -$ K}
\tablenotetext{e}{Intrinsic luminosity, adopting L$_{\odot} = 3.86e26$ W
then:
$ log_{10}L/L_{\odot} = log_{10}(flux) -2\times log_{10}\pi + 7.491 $}
\enddata
\end{deluxetable}

\newpage

\hoffset=-.65truein
\begin{deluxetable}{lcrrrc}
\tablenum{5}
\tablewidth{0pt}
\tablecaption{Derived Parameters for the Sample }
\tablehead{
\colhead{Spectral} & \colhead{Name} & \multicolumn{3}{c}{By Spectral Synthesis} & \colhead{By Structural Model}\nl
\colhead{Type}  &\colhead{}  &\colhead{$T_{\rm eff}$~K/$\log(g)$/[m/H]} & 
\multicolumn{2}{c}{Diameter $10^8~$m}  &\colhead{$T_{\rm eff}$~K} \nl
\colhead{}  &\colhead{} &\colhead{} & \colhead{by Scaling} & \colhead{by L,$T_{\rm eff}$}& \colhead{} \nl
}
\startdata
M5.5 & LHS 3406  &  2300/~6.0/\phs 0.0 & 2.38 & 2.42 & 2550--2700\nl
M6.5 & LHS 2930 & 2300/~6.0/\phs 0.0 & 2.09 & 2.14 & 2400--2650\nl
M7 & LHS 3003 & 2200/~6.0/\phs 0.0 & 2.20 & 2.20 & 2400--2650\nl
M7 & LHS 429 &  2300/~6.0/\phs 0.0 & 2.13 & 2.16& 2450--2650\nl
M8.5 & TVLM 513-46546 & 2200/~6.0/\phs 0.0 & 1.45 & 1.44 & 2000--2250  \nl
M9 & LP 944-20  & 2100/~6.0/\phs 0.0 & 1.43 & 1.38 & 1850--2100 \nl
M9 & LHS 2065 & 2100/~6.0/\phs 0.0 & 1.97 & 1.90 & 2150--2400\nl
M9.5 & BRI 0021-0214 & 2100/~6.0/\phs 0.0 & 2.12 & 2.03 & 2200--2450\nl
L0 & 2MASP J0345432$+$254023(SB?)  & 2000/~6.0/\phs 0.0 & 1.92 & 1.87 & 2050--2350 \nl
L2 & Kelu-1 & 1900/~5.5/\phs 0.0 & 2.10 & 2.14 & 2100--2350\nl
L3 & Denis-P J 1058$-$1548 & 1900/~5.5/\phs 0.0 & 1.28 & 1.32 & 1700-1950\nl
L4 & GD 165B& 1900/~5.5/\phs 0.0 & 1.14 & 1.20 & 1650--1850\nl
L5 & SDSS 0539 &  1900/~5.5/\phs 0.0 & \nodata & \nodata & \nodata\nl
L5 & Denis-P J 1228$-$1547AB & 1800/~5.0/$-$0.5 & 1.42 & 1.44 & 1450--1650\tablenotemark{a}\nl
L7 & Denis-P J 0205$-$1159AB & 1900/~5.5/\phs 0.0 & 1.11 & 1.20& 1400--1600\tablenotemark{a} \nl
\nl
\tablenotetext{a}{Assuming equal contribution to luminosity from each component of binary}
\enddata
\end{deluxetable}

\end{document}